\documentclass[useAMS,usenatbib]{mn2e}
\usepackage{amsmath,astrojournals,fleqn,graphicx,txfonts}

\title{The velocity dispersion and mass profile of the Milky Way}

\author[W.~Dehnen, D.~E.~McLaughlin and J.~Sachania]
{
  Walter Dehnen%
\thanks{Email: walter.dehnen@astro.le.ac.uk},
  Dean E.~McLaughlin,
  and Jalpesh Sachania\\
  Department of Physics \& Astronomy,
  University of Leicester,
  Leicester, LE1~7RH
}

\date{Accepted 2006 March 30.
      Received 2006 February 17;
      in original form 2006 January 24
      }

\pagerange{\pageref{firstpage}--\pageref{lastpage}}
\pubyear{2005}

\begin{document}

\maketitle
\label{firstpage}

\begin{abstract}
  We re-analyse the velocity-dispersion profile $\sigma(r)$ at radii
  $r>10\,$kpc in the Galactic stellar halo, recently derived by
  \citet{BattagliaEtal2005}, who concluded that, for a constant velocity
  anisotropy of the tracers, these data rule out a flat circular-speed curve
  for the Milky Way. However, we demonstrate that if one makes the natural
  assumption that the tracer density is truncated at
  $r_{\mathrm{t}}\ga160\,$kpc and falls off significantly more steeply than
  $r^{-3.5}$ at $r\ga80\,$kpc, then these data are consistent with a flat
  circular-speed curve and a constant velocity anisotropy comparable to that
  observed for halo stars in the Solar neighbourhood. We also consider a more
  detailed mass model with an exponential stellar disc and an extended
  non-isothermal dark-matter halo. In this two-component model, the Milky
  Way's virial radius and mass are $r_{\mathrm{vir}} \simeq 200\,$kpc and
  $M_{\mathrm{vir}} \simeq 1.5 \times 10^{12} \, {\mathrm{M}_\odot}$. Still
  assuming the tracers' velocity anisotropy to be constant (at
  $\beta\simeq0.5$) we again find good agreement with the observed
  $\sigma(r)$, so long as the tracer density is truncated near the virial
  radius. These data by themselves do not allow to differentiate between
  different dark-halo or total-mass models for the Milky Way, nor between
  different velocity-anisotropy profiles for the tracers.
\end{abstract}
\section{Introduction}
\label{sec:intro}
The case for dark-matter haloes around spiral galaxies is predominantly based
on the flatness of the rotation curves observed for gas out to large radii.
For large elliptical galaxies at the centres of clusters, similarly strong
evidence comes from X-ray haloes and from the kinematics of stellar tracers
such as globular clusters out to a few effective radii
\citep[e.g.,][]{CoteEtal2001,CoteEtal2003,RichtlerEtal2004}.  For more
isolated, `field' ellipticals, however---or even those off-centre within
clusters---the picture is somewhat less clear.  For example, the projected
velocity-dispersion profiles, $\sigma(r)$, of planetary nebulae in four such
galaxies have been measured to several effective radii by
\citet{MendezEtal2001} and \citet{RomanowskyEtal2003}.  Although these data
probe regions that are expected to be dynamically dominated by dark matter,
the dispersion profiles are found to decrease outwards in a manner consistent
with the assumption that light traces mass, i.e., that there are no extended
dark haloes. However, unlike rotation curves, $\sigma(r)$ profiles do not
directly measure the total mass distribution, but are also affected by the
density and velocity-anisotropy profiles of the tracers themselves. The
importance of velocity anisotropy particularly has been emphasised by
\citet{MamonLokas2005} and \citet{DekelEtal2005}, who showed that falling
dispersion profiles for elliptical galaxies are perfectly consistent with
massive dark-matter haloes, in contrast to the original conclusion by
\citet{RomanowskyEtal2003}.

\citet{BattagliaEtal2005} have recently derived a velocity-dispersion profile
for tracers at large radii in the Galactic stellar halo (globular clusters,
horizontal-branch and red-giant stars, and dSph satellites), finding a falling
$\sigma(r)$ much like in the isolated ellipticals. In analysing these data,
they carefully consider the effects of velocity anisotropy and conclude that
(1) a strictly flat circular-speed curve, corresponding to an isothermal
\emph{total} mass profile for our Galaxy, can be ruled out if the tracers'
velocity anisotropy is spatially constant, and (2) a massive dark halo is
consistent with the data only if its density falls off rapidly beyond
$r\ga100\,$kpc and/or the stellar tracers are on predominantly near-circular
orbits at these radii.

Underlying this argument by \citeauthor{BattagliaEtal2005} is an assumption
that the stellar tracers follow a power-law density profile, $\rho(r)\propto
r^{-3.5}$ all the way out. In their Appendix B, these authors briefly
addressed some possible consequences of relaxing this assumption, if the
tracer velocity-anisotropy profile is assigned one specific form. However,
they did not pursue any detailed modelling with the tracer $\rho(r)$ allowed
to deviate from a pure power law. The purpose of this paper is to
develop such models. We find that, since the velocity data from
\citeauthor{BattagliaEtal2005} reach more than halfway to the virial radius of
the Galaxy---probing regions where the stellar density is unconstrained
observationally, and even approaching what might be viewed as a natural `edge'
to the stellar halo---our ignorance of the detailed density distribution of
the far stellar halo is at least as important as uncertainties in the tracers'
velocity anisotropy when attempting to use the observed $\sigma(r)$ to infer
anything about the mass profile of the Galaxy.

Before considering the Milky Way data specifically, let us first review the
factors affecting the velocity dispersion of a tracer population. We assume
dynamical equilibrium and spherical symmetry, denote the tracers' density
profile by $\rho(r)$, and characterise the total galactic mass distribution by
its circular-speed curve $V_{\mathrm{c}}^2(r)=GM_{\mathrm{tot}}(r)/r$. Then
the radial component $\sigma_r(r)$ of the tracer population's velocity
dispersion must satisfy the Jeans equation
\begin{equation}
  \label{eq:jeans}
  \frac{\mathrm{d}}{\mathrm{d}r} \left(\rho\sigma_r^2\right) +
  \frac{2\beta}{r}\,\rho\sigma_r^2 = - \rho\,\frac{V_{\mathrm{c}}^2}{r} \ ,
\end{equation}
where $\beta(r)=1-\sigma_\theta^2/\sigma_r^2$ is the usual velocity anisotropy
parameter of the tracers ($\beta=0$ for isotropy, $0<\beta\le 1$ for radial
anisotropy, and $\beta<0$ for a tangentially biased velocity distribution).

In the simple situation of constant anisotropy $\beta$, and power laws for
both the tracer density profile and the galaxy's circular-speed curve,
$\rho\propto r^{-\gamma}$ and $V_{\mathrm{c}}\propto r^\alpha$, the Jeans
equation requires that
\begin{equation}
  \label{eq:sigma:power}
    \sigma_r^2 = \frac{1}{\gamma-2\beta-2\alpha}\,V_{\mathrm{c}}^2(r)
    \quad\text{and}\quad
    \sigma_\theta^2 = 
    \frac{1-\beta}{\gamma-2\beta-2\alpha}\,V_{\mathrm{c}}^2(r) \ ,
\end{equation}
which reduces to equation (B2) of \citet{BattagliaEtal2005} in the special
case $\alpha=0$ ($V_{\mathrm{c}}=\mathrm{constant}$). Thus, both components of
$\sigma$ are proportional to $V_{\mathrm{c}}$, and they fall with radius if
and only if the circular speed does. More realistically however, $\beta$
and/or $\gamma\equiv -\mathrm{d} \ln \rho/\mathrm{d}\ln r$ will be functions
of radius; but we expect that at relatively large radii these will vary slowly
enough for equation (\ref{eq:sigma:power}) to be still approximately valid and
provide some heuristic insight into the general behaviour of $\sigma(r)$.

First, then, let $\beta(r)$ decrease with radius, so that the tracer velocity
distribution becomes more tangentially biased at larger radii. In this case,
the denominators in equation (\ref{eq:sigma:power}) increase, so that
$\sigma_r(r)$ falls relative to $V_{\mathrm{c}}(r)$. Thus
\citet{BattagliaEtal2005}, working with strictly constant $\gamma$, conclude
that a rather strongly decreasing anisotropy profile is necessary to reconcile
their observed falling $\sigma_r(r)$ profile with a standard
\citet*[][`NFW']{NavarroFrenkWhite1996} model for the mass distribution of the
Milky Way.

For $\gamma\ga2(1+\alpha)$ (which holds in the Galactic halo and the outskirts
of ellipticals), the effect on $\sigma_\theta$ is dominated by the
\emph{numerator} in equation~(\ref{eq:sigma:power}), such that a decreasing
$\beta(r)$ results in $\sigma_\theta(r)$ rising relative to
$V_{\mathrm{c}}(r)$. The effect on the projected line-of-sight velocity
dispersion is dominated by $\sigma_\theta$ (but complicated by the projection)
and results in the well-known degeneracy between the velocity anisotropy and
the mass profile, which has plagued the interpretation of the
velocity-dispersion profiles of ellipticals \citep[for a recent exploration of
this problem, see][]{MamonLokas2005}.

Second, suppose instead that $\gamma(r)$ increases with radius
(corresponding to $\rho(r)$ decaying faster than a power law) while $\beta$
remains constant. In this case, equation~(\ref{eq:sigma:power}) tells us that
both $\sigma_r(r)$ and $\sigma_\theta(r)$ fall relative to
$V_{\mathrm{c}}(r)$. In fact, if---for example---the tracer density decays
exponentially, the anisotropy is fixed at $\beta=1/2$, and the galaxy circular
speed is constant ($\alpha=0$), then we find from an \emph{exact} solution of
the Jeans equation~(\ref{eq:jeans}) that $\sigma_r\propto r^{-1/2}$: the
velocity dispersion declines like that for a power-law $\rho(r)$ in a
Keplerian potential, even though the rotation curve is flat!

An even more drastic behaviour results from a tracer population which is
truncated at a finite radius $r_{\mathrm{t}}$. In this case, $\gamma\to\infty$
as $r\to r_{\mathrm{t}}$ and from equation~(\ref{eq:sigma:power}) we expect
both components of the velocity dispersion to vanish in this limit.  Thus, the
falling $\sigma_r(r)$ observed for the stellar halo of the Milky Way may still
be consistent with a roughly flat circular-speed curve, if the tracer
population effectively dies out at some large radius. This is the possibility
that we pursue in more detail in Section~\ref{sec:MW}. Note that for
elliptical galaxies the tracer density is known, so that this problem does not
arise in the interpretation of their velocity dispersions.

\section{Modelling the Milky Way}
\label{sec:MW}

\subsection{A geometrical correction factor}
\label{sec:geocorr}

The data collected by \citet{BattagliaEtal2005} comprise heliocentric radial
velocities corrected for the Solar motion relative to the Galactic centre,
yielding velocities, $v_{\mathrm{GSR}}$, in a Galactic standard of rest.
However, these velocities must still be referred to the Galactic centre before
comparing to any model. Thus, using spherical polar coordinates with the
centre of the Milky Way at the origin, and the Sun on the $z$-axis at distance
$R_0=8\,$kpc, we express the instantaneous distance $d$ from an object to the
Sun as
\begin{equation}
  \label{eq:app:dist}
  d^2 = r^2 + R_0^2 - 2 r R_0 \cos\theta \ .
\end{equation}
The time derivative of this is $\dot{d}=v_{\mathrm{GSR}} = a_r v_r +
a_\theta v_\theta$, where
\begin{equation}
  \label{eq:app:AB}
  a_r = (r-R_0\cos\theta)/d
     \quad\text{and}\quad
  a_\theta = (R_0\sin\theta)/d \ .
\end{equation}
The velocity dispersion at a given Galactocentric radius $r$ is estimated as
the rms velocity averaged over all solid angles:
\begin{equation}
  \label{eq:app:sigma:gsr}
  \sigma_{\mathrm{GSR}}^2 =
  \langle a_r^2 \rangle\, \sigma_r^2 +
  \langle a_\theta^2 \rangle\, \sigma_\theta^2,
\end{equation}
where $\langle.\rangle$ denotes the averaging over the sphere. Using
$a_r^2+a_\theta^2=1$ and $\sigma_\theta^2=(1-\beta)\sigma_r^2$, we thus obtain
\begin{equation}
  \label{eq:sigma:corr}
  \sigma_{\mathrm{GSR}}(r) = \sigma_r\sqrt{1-\beta(r)\, A(r)}
\end{equation}
with $A(r)\equiv \langle a_\theta^2 \rangle$ a monotonically declining
function of radius:
\begin{equation}
  \label{eq:Atheta}
  A(r) = \frac{r^2+R_0^2}{4r^2}
  - \frac{(r^2-R_0^2)^2}{8r^3R_0} \ln\left|\frac{r+R_0}{r-R_0}\right|.
\end{equation}
The correction factor in equation~(\ref{eq:sigma:corr}) becomes unity for an
isotropic velocity distribution ($\beta=0$), since in this case the velocity
dispersion is independent of direction. For non-zero $\beta$, the correction
is greatest for small $r$, where $A(r)$ becomes largest ($A=1/2$ at $r=R_0$
and $A=2/3$ at $r=0$), while for $r\gg R_0$, $A\ll 1$ and
$\sigma_{\mathrm{GSR}}\approx\sigma_r$.

Unfortunately, \citet{BattagliaEtal2005} use an incorrect formula to convert
between $\sigma_r$ and $\sigma_{\mathrm{GSR}}$ [their equation~(3) corresponds
to (\ref{eq:sigma:corr}) above], which gives a correction factor larger than
unity for $\beta=0$. This significantly affects the predicted
$\sigma_{\mathrm{GSR}}$ only at radii $r\la 30\,$kpc. However, this is where
the uncertainties in the observed $\sigma_{\mathrm{GSR}}$ are smallest, 
resulting in substantially erroneous $\chi^2$ for any model fits.

\subsection{Simple models with truncated tracer populations}
\label{sec:trunc}

\begin{figure}
  \centerline{\hfil
    \resizebox{82mm}{!}{\includegraphics{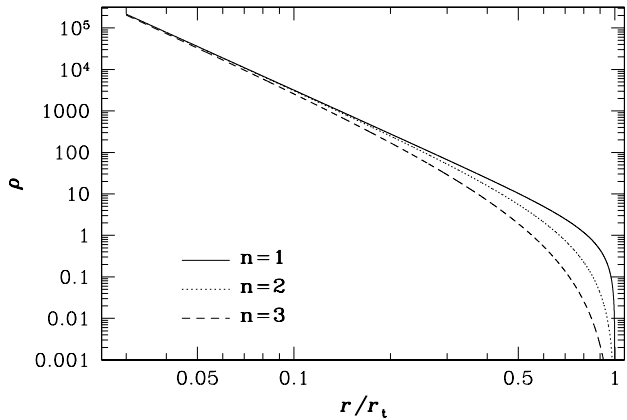}}
  }
  \caption{
    \label{fig:rho:trunc}
    Truncated density profiles of equation~(\ref{eq:rho:trunc}) with
    $\gamma=3.5$ and $n=1$ (solid), $n=2$ (dotted), or $n=3$ (dashed).  }
\end{figure}

Let us now employ some simple, fully analytic models of truncated density
profiles to demonstrate that the Milky Way $\sigma_{\mathrm{GSR}}$ data are
consistent with a flat circular-speed curve, i.e., an isothermal total-mass
distribution. We assume that the density of the tracer population, outside of
some (small) core radius, is given by
\begin{equation}
  \label{eq:rho:trunc}
  \rho(r) \propto 
  \begin{cases}
    \left(r^{-\gamma/n} -  r_{\mathrm{t}}^{-\gamma/n}\right)^n
      & \text{for $r<r_{\mathrm{t}}$,} \\
    0 & \text{for $r\ge r_{\mathrm{t}}$,}
  \end{cases}
\end{equation}
which is plotted in Fig.~\ref{fig:rho:trunc} for $\gamma=3.5$ and $n=1$, 2, or
3. At small radii, say $r\la0.5r_{\mathrm{t}}$, the profiles in this graph are
hardly distinguishable from a pure power-law $\rho\propto r^{-3.5}$, which
matches the density of the Galactic stellar halo out to $r\sim50\,$kpc
\citep{MorrisonEtal2000,YannyEtal2000}. The abruptness of the truncation
towards larger radii depends on the parameter $n$. We have used the online
database of \citet{Harris1996} to verify that, for $r_{\mathrm{t}}\ga
150\,$kpc, a model with $\gamma=3.5$ and $n=2$ provides a good description of
the density of Galactic globular clusters at $r\ga3\,$kpc. This is thus the
model we use for all applications to data in this paper.

\begin{figure}
  \centerline{\resizebox{82mm}{!}{\includegraphics{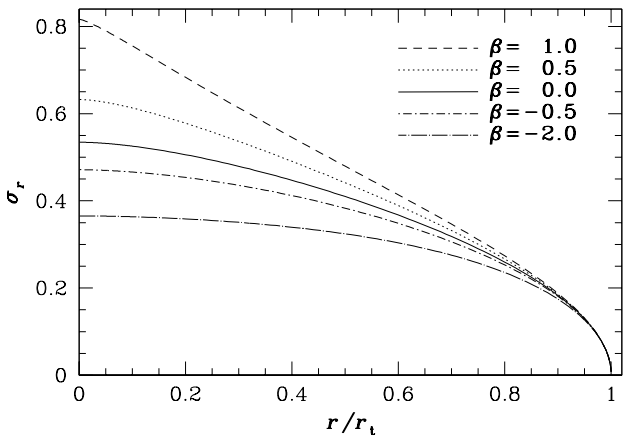}}\hfil}
  \caption[]{
    \label{fig:sig:trunc}
    Radial velocity-dispersion profile for a tracer population with constant
    anisotropy $\beta$ and truncated density~(\ref{eq:rho:trunc}) with
    $\gamma=3.5$ and $n=2$, embedded in a gravitational potential with
    constant circular speed ($\alpha=0$). For a slightly falling
    circular-speed curve, such as predicted for dark-matter haloes, $\sigma_r$
    falls off slightly faster.}
\end{figure}

Given equation (\ref{eq:rho:trunc}), a constant anisotropy $\beta$ for the
tracers, and a power-law circular-speed curve, $V_{\mathrm{c}}(r)=V_0
(r/r_{\mathrm{t}})^{\alpha}$, the radial component of the velocity dispersion
can be obtained analytically from the Jeans equation (\ref{eq:jeans}) for
integer $n$:
\begin{equation}
  \label{eq:sigmar:trunc}
  \sigma_r^2 = \frac{V_0^2}
  {(x^{-\gamma/n}-1)^n x^{2\beta}}\,
  \sum_{k=0}^n \binom{n}{k}\,(-1)^{n-k}\,
  h\left(\frac{\gamma k}{n}-2\alpha-2\beta;\,x\right)
\end{equation}
with $x\equiv r/r_{\mathrm{t}}$ and
\begin{equation}
  h(q;x)\equiv\int_x^1 \frac{\mathrm{d}u}{u^{q+1}} = 
  \begin{cases} 
    \;- \ln x & \text{for $q=0$}, \\[0.5ex]
    \displaystyle
    \;\frac{1}{q}(x^{-q}-1) & \text{for $q\neq0$}.
  \end{cases}
\end{equation}
In the limit $r\to r_{\mathrm{t}}$, $\sigma_r\to0$, as the simple
considerations of Section~\ref{sec:intro} already implied it must. (In this
limit equation~(\ref{eq:sigmar:trunc}) is numerically unstable against
truncation errors, but instead one may use
\begin{equation}
  \label{eq:sigmar:trunc:limit}
  \frac{\sigma_r^2(r)}{V_{\mathrm{c}}^2(r)} = \frac{\epsilon}{n+1} 
  \left[1+\left( n+1+2(\alpha+\beta)-\frac{\gamma+n}{2}\right)
    \frac{\epsilon}{n+2}
    + \mathcal{O}(\epsilon^2)\right]
\end{equation}
with $\epsilon=1-x>0$.)

In Fig.~\ref{fig:sig:trunc}, we plot $\sigma_r(r)$ for a flat rotation curve
($\alpha=0$) and $\gamma=3.5$, $n=2$ (appropriate to the situation in the
Milky Way stellar halo), and for various $\beta$. Evidently, the velocity
dispersion declines over much the whole range of radii, even though the
density deviates from the power-law form only for large radii.  The physical
reason is that in order to achieve a truncation in density, orbits that reach
near $r_{\mathrm{t}}$ must be de-populated relative to a non-truncated model.
These orbits, however, would have contributed at smaller radii with rather
large $v_r$. Note that all the models presented in Fig.~\ref{fig:sig:trunc}
are physical in the sense that they possess a non-negative distribution
function of the form $f(E,L)=L^{-2\beta}g(E)$.

\begin{figure}
  \centerline{\resizebox{82mm}{!}{\includegraphics{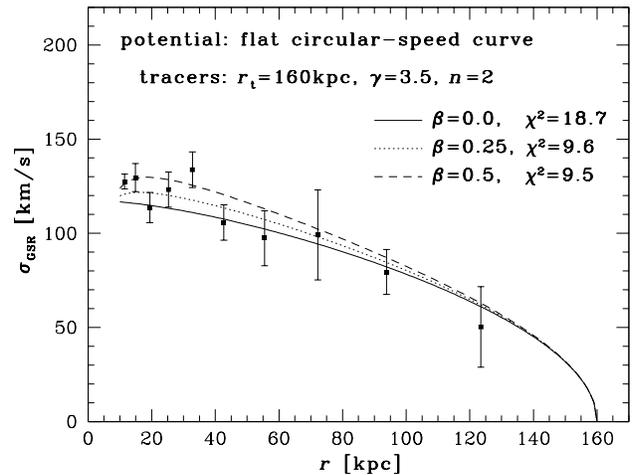}}\hfil}
  \caption[]{
    \label{fig:trunc:MW}
    Comparison of truncated power-law models embedded in a flat circular-speed
    curve of $V_{\mathrm{c}}=220\,$km\,s$^{-1}$ to the
    \citet{BattagliaEtal2005} data for the Milky Way. Slightly better fits
    could be achieved with smaller truncation radius $ r_{\mathrm{t}}$ or a
    slightly falling circular-speed curve.  }
\end{figure}
In Fig.~\ref{fig:trunc:MW}, we compare these simple models (after applying the
geometric correction of Section~\ref{sec:geocorr}) to the
$\sigma_{\mathrm{GSR}}(r)$ data of \citet{BattagliaEtal2005}. The tracer
density again has $\gamma=3.5$ and $n=2$, and we set the truncation radius to
$ r_{\mathrm{t}}=160\,$kpc. We assume a constant circular speed of
$V_{\mathrm{c}}=220\,$km\,s$^{-1}$ and constant $\beta$ of 0 (isotropic),
0.25, or 0.5 (radially anisotropic). All these models correctly reproduce the
decline of $\sigma$ towards larger radii.

At small radii ($r\la40\,$kpc), the isotropic model (solid curve) cannot
account for the observed dispersion, but falls significantly below.  Since at
these radii $\gamma\approx3.5$ is observationally secure, we have from
equation~(\ref{eq:sigma:power}) that $\sigma_r \approx
V_{\mathrm{c}}/\sqrt{3.5}=0.53V_{\mathrm{c}}$ for any isotropic model if
$V_{\mathrm{c}}(r)$ is near-flat just outside the solar circle. A possible
explanation of the observed high $\sigma_{\mathrm{GSR}}$ for a tracer
population with isotropic velocities requires a circular-speed curve which
rises like $V_{\mathrm{c}}\propto r^{0.1}$ out to $r\sim30\,$kpc (where then
$V_{\mathrm{c}}=250\,$km\,s$^{-1}$). Alternatively and more reasonably, the
data are consistent with a near constant $V_{\mathrm{c}}=220\,$km\,s$^{-1}$ if
the tracer velocities at $r\la40\,$kpc are slightly radially biased with
$\beta=0.25-0.5$, comparable to what has been derived for halo stars in the
Solar neighbourhood \citep{ChibaYoshii1998,Gould2004E}.

\subsection{More sophisticated models}
\label{sec:soph}

The above analysis shows that the \citet{BattagliaEtal2005} data \emph{are} in
fact consistent with constant velocity anisotropy for the tracers and an
isothermal total mass distribution for the Milky Way. We can further ask
whether the same data are consistent with a more detailed model explicitly
including both a stellar disc---which is well known to dominate the total mass
inside the Solar circle \citep[e.g.,][]{DehnenBinney1998}, and still
contributes $\ga10\%$ to the mass at 50\,kpc---and an extended dark-matter
halo which is (as per cosmological $N$-body simulations) non-isothermal.

We first specify an exponential stellar disc with rotation curve
\begin{equation}
  \label{eq:vcdisc}
  V_{\mathrm{c,d}}^2(r) = \frac{GM_{\mathrm{d}}(r)}{r} = 
       \frac{GM_{\mathrm{d,tot}}}{r}\,
       \left[1 - \left(1+\frac{r}{r_{\mathrm{d}}}\right)\,
       \exp\left(-\frac{r}{r_{\mathrm{d}}}\right)\right] \ ,
\end{equation}
where $M_{\mathrm{d,tot}}=5.8\times10^{10}\,{\mathrm{M}_\odot}$ and
$r_{\mathrm{d}}=2.4\,$kpc according to \citet{DehnenBinney1998}.

Second, we take a dark-matter halo from the family of models developed by
\citet{DehnenMcLaughlin2005}. Specifically, we assume that the halo follows the
scaling $\rho_{\mathrm{h}}/\sigma_{r,\mathrm{h}}^3 \propto r^{-35/18}$
(consistent with simulations) and has an isotropic velocity distribution at
its centre (but may be anisotropic elsewhere). Such a halo has a density cusp
$\rho_{\mathrm{h}}\to r^{-7/9}$ in the limit $r\to 0$, and
$\rho_{\mathrm{h}}\to r^{-31/9}$ as $r\to\infty$.  Thus, it has a finite total
mass, and its circular-speed curve is
\begin{equation}
  \label{eq:vchalo}
  V_{\mathrm{c,h}}^2(r) = \frac{GM_{\mathrm{h}}(r)}{r} =
  \frac{GM_{\mathrm{h,tot}}}{r}\,
  \left[\frac{r^{4/9}}{r^{4/9}+r_0^{4/9}}\right]^5,
\end{equation}
from equation (20f) of \citet{DehnenMcLaughlin2005}.

To fix the halo parameters, we specify the value of the \emph{total} circular
speed, $V_{\mathrm{c}}^2=V_{\mathrm{c}, d}^2+V_{\mathrm{c,h}}^2$, at two
Galactocentric radii: at the position of the Sun, $r=8\,$kpc, we require
$V_{\mathrm{c}}=220\,$km\,s$^{-1}$, while at $r=50\,$kpc we adopt
$V_{\mathrm{c}}=205\,$km\,s$^{-1}$.  This latter value corresponds to a total
mass of $M_{\mathrm{d}} + M_{\mathrm{h}} = 4.9 \times 10^{11} \,
{\mathrm{M}_\odot}$ inside $r=50\,$kpc, chosen in order to agree with the
analysis of \citet[][see also
\citealt{RohlfsKreitschmann1988}]{Kochanek1996}.  These two constraints then
imply $M_{\mathrm{h,tot}} = 1.10\times 10^{13}\,{\mathrm{M}_\odot}$ and
$r_0=40.5\,$kpc.

\begin{figure}
  \centerline{\resizebox{82mm}{!}{\includegraphics{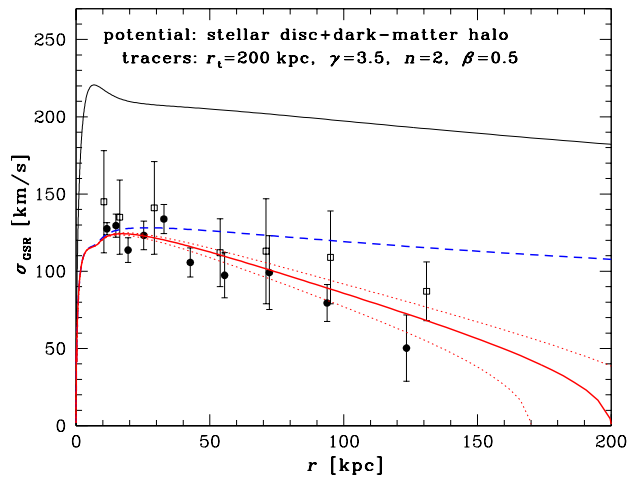}}\hfil}
  \caption[]{
    \label{fig:dischalo}
    Two-component mass model for the Milky Way, and the velocity-dispersion
    profile for a tracer population in the stellar halo. The thin, solid ({\it
      black}) curve is the total circular-speed curve $V_{\mathrm{c}}(r)$ of
    the model (see eqs.~[\ref{eq:vcdisc}] and [\ref{eq:vchalo}]). The bold,
    solid (\emph{red}) curve running through the data is the predicted GSR
    velocity-dispersion profile for a tracer population with a density profile
    truncated at $ r_{\mathrm{t}}=200\,$kpc (see text for details) and with a
    spatially constant velocity anisotropy $\beta=0.5$. The thinner, dotted
    (\emph{red}) curves bracketing this are the predicted
    $\sigma_{\mathrm{GSR}}$ for $\beta=0.5$ still, but with truncation radii $
    r_{\mathrm{t}}=170\,$kpc and $ r_{\mathrm{t}}=230\,$kpc. The bold, dashed
    (\emph{blue}) curve above these is the $\sigma_{\mathrm{GSR}}$ that would
    obtain with $\beta=0.5$ and $\rho(r)\propto r^{-3.5}$ at all radii (no
    truncation).  Data points are from \citet{BattagliaEtal2005} (filled
    circles) and from \citet{Harris2001} (open squares).  }
\end{figure}

The total mass $M_{\mathrm{h,tot}}$ here is obtained formally by integrating
the dark-matter density profile to infinity, but physically more meaningful is
the halo mass within finite radii corresponding to specific over-densities
relative to the critical $\rho_{\mathrm{c}}=3H_0^2/8\pi G$. First, in a
`concordance' $\Lambda$-CDM cosmology with $\Omega_m=0.3$ and
$\Omega_\Lambda=0.7$, the virial radius of a halo is that within which the
average density $3M_{\mathrm{h}}(r_{\mathrm{vir}})/4\pi r_{\mathrm{vir}}^3$ is
equal to $337\rho_{\mathrm{c}}$ \citep[e.g.,][]{BullockEtal2001a}. With
$H_0=70\,$km\,s$^{-1}$\,Mpc$^{-1}$, $r_{\mathrm{vir}}\simeq200\,$kpc and
$M_{\mathrm{vir}}\simeq1.5\times10^{12}\,{\mathrm{M}_\odot}$ for our model,
both of which are reasonable for an $L_*$ galaxy like the Milky Way.
Alternatively, haloes are commonly measured by the radius $r_{200}$ within
which the average density is $200\rho_{\mathrm{c}}$. In our case, $r_{200}
\simeq 250\,$kpc and $M_{200} \simeq 1.75 \times 10^{12}\,
{\mathrm{M}_\odot}$. For comparison with numbers given by
\citet{BattagliaEtal2005}, the total mass inside 120\,kpc is
$M(r\le120\,{\mathrm{kpc}})=1.05\times10^{12}\,{\mathrm{M}_\odot}$.

A scale of interest in connection with numerical simulations of
dark-matter haloes is the radius $r_{-2}$ at which the local logarithmic slope
of the density profile is $\mathrm{d}\ln \, \rho_{\mathrm{h}}/\mathrm{d} \ln
\,r=-2$. In our model, $r_{-2}=(11/13)^{9/4}\,r_0=27.8\,$kpc, and the ratio
$r_{200}/r_{-2} = 9$ is nicely consistent with the values found by
\citet{NavarroEtal2004} for simulated haloes with masses in the range
$M_{200}=1-2\times10^{12}\,h^{-1}{\mathrm{M}_\odot}$.

To predict the kinematics of tracers in the stellar halo, we use the density
model for the tracers that has already been employed in
section~\ref{sec:trunc}, i.e., equation~(\ref{eq:rho:trunc}) with $\gamma=3.5$
and $n=2$, although we set the truncation radius $
r_{\mathrm{t}}=r_{\mathrm{vir}}= 200\,$kpc, somewhat larger than the value
used in Figure \ref{fig:trunc:MW}.  We then solve the Jeans
equation~(\ref{eq:jeans}) for an assumed spatially constant velocity
anisotropy $\beta$, and apply the geometric correction in
equation~(\ref{eq:sigma:corr}). This yields a model
$\sigma_{\mathrm{GSR}}(r)$, which is compared to the \citet{BattagliaEtal2005}
data to compute $\chi^2$. The minimum $\chi^2=9.5$ (for 10 data points) is
achieved with $\beta=0.5$. As was mentioned above, this slight radial bias is
consistent with observations of halo stars in the Solar neighbourhood
\citep{ChibaYoshii1998, Gould2004E}.

Figure \ref{fig:dischalo} shows our best-fit $\sigma_{\mathrm{GSR}}$ profile
against the \citeauthor{BattagliaEtal2005} data, plus another estimate of the
stellar-halo velocity-dispersion profile from \citet{Harris2001}.  The latter
is also constructed from velocity data for globular clusters, RR~Lyrae stars,
and dwarf spheroidals, and so is not independent of the
\citeauthor{BattagliaEtal2005} profile; thus, we have not used it in
determining $\chi^2$ for our models. However, this alternate profile serves to
confirm the overall sense of the \citeauthor{BattagliaEtal2005} results (and
to emphasise their uncertainty at the largest radii).  We have also plotted in
Figure \ref{fig:dischalo} alternate models in which the tracers still have
constant $\beta=0.5$ but are truncated at $r_{\mathrm{t}}=170$ or 230\,kpc.
The bold, dashed curve which declines only gradually towards large radius is
the velocity-dispersion profile assuming the tracer density profile to be an
untruncated pure power law $\rho\propto r^{-3.5}$. This is the assumption made
by \citet{BattagliaEtal2005} in their modelling, and it clearly has a
dramatic---even dominant---influence on the anisotropy profiles they require
in order to fit the observed $\sigma_{\mathrm{GSR}}(r)$.

\section{Discussion}
\label{sec:disc}
Comparing Figures \ref{fig:trunc:MW} and \ref{fig:dischalo}, it is clear that
in the latter we are able to describe the observed $\sigma_{\mathrm{GSR}}(r)$
profile with a larger assumed truncation radius for the stellar-halo tracers
than in the former, and also that a constant $\beta=0.5$ predicts slightly
lower velocity dispersions at small radii in our (disc+halo) mass model than
in the constant-$V_{\mathrm{c}}$ model.  These points simply reflect that the
total circular speed given by equations (\ref{eq:vcdisc}) and
(\ref{eq:vchalo}) decreases monotonically with radius for $r>6.6\,$kpc (which
adds to the effect of a truncated tracer density in driving the decline of
$\sigma_{\mathrm{GSR}}$), and in fact is less than 220\,km\,s$^{-1}$ at all
radii covered by the \citet{BattagliaEtal2005} or \citet{Harris2001} data (so
for fixed $\gamma$, a slightly higher $\beta$ is required to give the same
normalisation to the tracers' $\sigma_r$, as can be seen from
eq.~[\ref{eq:sigma:power}]).

Even so, all of the anisotropic models illustrated here fit the data equally
well, and thus the tracer velocity-dispersion profile alone {\it cannot} be
used to distinguish between different (reasonable) models for the total mass
distribution in the Milky Way.  Conversely, beyond concluding simply that the
observed $\sigma_{\mathrm{GSR}}(r)$ profile is consistent with a tracer
density profile that falls steeply at $r\ga80\,$kpc, we cannot use these
kinematics to infer anything more detailed about the behaviour of $\rho(r)$
for the far stellar halo.

Nor can strong constraints be placed on the velocity anisotropy of the tracer
population. In the context of a spatially constant $\beta$, as we have
assumed, all that can be said with any confidence is that a slight radial bias
($\beta\sim0.2-0.6$, depending on the Galactic mass model) is required to
explain the amplitude of $\sigma_{\mathrm{GSR}}$ at $r\la40\,$kpc. If the
assumed form of the total mass distribution deviates significantly from the
models with constant or slowly-varying $V_{\mathrm{c}}(r)$ that we have
explored, then a variety of $\beta(r)$ behaviour is likely allowed by the
data.

Clearly, untangling the degeneracies between the total $M(r)$ and tracer
$\rho(r)$ and $\beta(r)$, to put quantitative limits on any one profile from
knowledge of the others, is a daunting task that will require much closer
attention to a wider variety of data beyond just $\sigma_{\mathrm{GSR}}(r)$.
For example, the models for $V_{\mathrm{c}}(r)$ that we have worked with here
are extremely simple and would surely require some modification in detail if
we attempted to take accurately into account the many other constraints on the
Galactic mass distribution. But more detailed modelling will also have to
allow for the fact that the tracers contributing to the
\citet{BattagliaEtal2005} and \citet{Harris2001} velocity-dispersion profiles
are very much a `mixed bag': the halo globular clusters, for instance, may not
have the same anisotropy profile as the field RR~Lyrae and red giants; and the
satellite dwarf spheroidals likely do not follow the same density profile as
the stars and globulars. Ideally, we need information on {\it separate}
$\sigma(r)$, $\rho(r)$, and $\beta(r)$ profiles for each of the different
tracer populations. Current data simply do not provide this.

\section{Summary}
\label{sec:sum}
We have demonstrated in this study that the falling velocity dispersion
found by \citet{BattagliaEtal2005} in the outer stellar halo of the Milky Way
is consistent with a constant (and reasonable) velocity anisotropy for the
stellar tracers, and either a perfectly flat circular-speed curve (i.e.\ an
isothermal total mass distribution) or a standard CDM dark-matter halo
combined with an exponential stellar disc. By contrast,
\citeauthor{BattagliaEtal2005} argued that their data were inconsistent with
$V_{\mathrm{c}}$ and $\beta$ both being strictly constant, and that they
required a strongly varying (and rather unusual) $\beta(r)$ profile to be made
compatible with the common \citet{NavarroFrenkWhite1996} model for a
dark-matter halo.%
\footnote{When \citet{BattagliaEtal2005} discuss `NFW' or `truncated flat' or
  `isothermal' haloes, they really model the \emph{total} Galactic mass
  distribution with those functions. But as we discussed in connection with
  Figure~\ref{fig:dischalo}, the stellar disc contributes significantly to the
  total circular speed out to tens of kpc, and the total $V_{\mathrm{c}}(r)$
  differs from that of the dark halo until \emph{very} large radii.}

As we have emphasised, the primary reason for the difference between our
conclusions and theirs is that we allow for the tracer density to die out at
$r\ga\,$160kpc, close to the Milky Way's virial radius, whereas
\citeauthor{BattagliaEtal2005} assume that $\rho\propto r^{-3.5}$, which is
valid for $r\la50\,$kpc, continues to hold for all larger radii.  An incorrect
transformation from modelled $\sigma_r(r)$ to observed
$\sigma_{\mathrm{GSR}}(r)$ on the part of \citeauthor{BattagliaEtal2005} (see
Section \ref{sec:geocorr}) has certainly also contributed to the discrepancy,
insofar as it affected the $\chi^2$ values of their detailed model fits.

As we discussed, these data by themselves are insufficient for differention
between different total-mass models for the Galaxy, nor between different
velocity-anisotropy profiles for the tracers. This is essentially because the
measured $\sigma(r)$ depends on both of the above (the well-known degeneracy
between velocity anisotropy and mass profile) and on the tracer's density
profile, which is little constrained observationally at $r\ga50\,$kpc.

\section*{Acknowledgements}
We thank Giuseppina Battaglia for providing us with the Milky Way
velocity-dispersion data points from her Figure 1, in electronic form.  DEM is
supported by a PPARC standard grant, and research in theoretical astrophysics
at the University of Leicester is also supported by a PPARC rolling grant.



\label{lastpage}
\end{document}